\documentclass[useAMS,usenatbib]{mn2e}
\usepackage[dvips]{graphicx,color}
\usepackage{txfonts}
\usepackage{amssymb}
\usepackage{indentfirst}
\usepackage{epsfig}

\newcommand{\MgXII}{\mbox{{Mg\,{\sevensize XII}}}}

\newcommand{\AlXIII}{\mbox{{Al\,{\sevensize XIII}}}}

\newcommand{\Lya}{\ensuremath{\hbox{Ly}\alpha~}}
\newcommand{\Lyb}{\ensuremath{\hbox{Ly}\beta~}}
\newcommand{\Ka}{\ensuremath{\hbox{K}\alpha~}}

\newcommand{\A}{\AA~}

\def\cha{{\it Chandra }}

\begin{document}
\title{On the emitting region of X-ray fluorescent lines around Compton-thick AGN} 

\author[J. Liu et al.]{Jiren Liu\thanks{E-mail: jirenliu@nao.cas.cn}  \\
	 $^{}$National Astronomical Observatories, 20A Datun Road, Beijing 100012, China\\
}

\date{}

\maketitle

\begin{abstract}

X-ray fluorescent lines are unique features of the reflection spectrum
of the torus when irradiated by the central AGN. 
Their intrinsic line width can be used to probe the line-emitting region.
Previous studies have focused on the Fe \Ka line at 6.4 keV, which is the most 
prominent fluorescent line. These studies, however, are limited by the 
spectral resolution of currently available instruments, the best of which is
$\sim1860$ km\ s$^{-1}$ afforded by the \cha High-Energy Grating (HEG).
The HEG spectral resolution is improved by a factor of 4 at 1.74 keV, where the 
Si \Ka line is located.
We measured the FWHM of the Si \Ka line for Circinus, Mrk 3, and NGC 1068, which
are $570\pm240$, $730\pm320$, and $320\pm280$ km\ s$^{-1}$, respectively.
They are $3-5$ times smaller than those measured with the Fe \Ka line previously.
It shows that the intrinsic widths 
of the Fe \Ka line are most likely to be over-estimated.
The measured widths of the Si \Ka line put the line-emitting region
outside the dust sublimation radius in these galaxies.
It indicates that for Compton-thick AGN, the X-ray fluorescence material are likely 
to be the same as the dusty torus emitting in the infrared.

\end{abstract}


\section{Introduction}

The obscuring torus around active galactic nuclei (AGN) is a cornerstone
of the unification scheme of AGN. In the traditional model the inclination of the
torus to the observer is supposed to be responsible for the major differences between 
type I and type II AGN \citep[e.g.][]{Ant93}. 
The location of torus was first considered to be on parsec scale, large enough to obscure the
broad line region (BLR) and small enough not to obscure the narrow line region \citep{KB88}.
This is confirmed by recent infrared interferometric observations of several nearby galaxies 
\citep[see][for a recent review]{Net15}.
The presence of the obscuring torus is also revealed by the Compton reflection component 
in the X-ray spectra of AGN, which is characterized by the fluorescent Fe \Ka line (6.4 keV).
The parsec-scale distance of the emitting region of the Fe \Ka line is evidenced by 
the stability of the Fe \Ka line \citep[e.g.][]{Bia12}. The exact location of the emitting region
can be estimated based on the intrinsic width of the X-ray fluorescent lines 
\citep[e.g.][and reference therein]{Shu11}. 

The measurement of the intrinsic line width, however, is limited by the spectral 
resolution of currently available instruments.
The \cha High Energy Transmission 
Grating Spectrometer \citep[HETGS,][]{Can05} provides a best spectral resolution of 0.012 \A 
(full width half maximum, FWHM) with its High Energy Grating (HEG), which
at 6.4 keV corresponds to $\sim1860$ km\ s$^{-1}$, very close to the measured mean FWHM
of the Fe line (2000 km\ s$^{-1}$) by \citet{Shu11}. 
We note that the HEG spectral resolution can be improved for lines at lower energies.
Therefore, the measurement of the line width
can be improved by using the other fluorescent lines, such as
the Si \Ka line at 1.74 eV, where the HEG spectral resolution is $\sim500$ km s$^{-1}$,
a factor of 4 times better than that at 6.4 keV.

The non-Fe fluorescent lines are generally weak, due to their small yields.
Nevertheless, they are still observable, especially for Compton-thick AGN, 
the intrinsic continua of which are heavily suppressed below 10 keV and the observed spectra
are dominated by the reflected component from the torus \citep[e.g.][]{Com04}.
For example, a series of \Ka fluorescent lines of
Si, S, Ar, Ca, Cr, and Mn are detected in M51, a low-luminosity Compton-thick AGN
\citep{Xu16}.

In this Letter we compile a sample of Compton-thick AGN observed with \cha
HETGS to measure their intrinsic width of the Si \Ka line and estimate the
location of the emitting region of the X-ray fluorescent lines.
The errors quoted are for 90\% confidence level.

\begin{table*}
	\caption{List of the Compton-thick AGN observed with \cha HETGS}
\begin{tabular}{lccccccccc}
\hline
\hline
Name & I(Si K$\alpha$)& $\sigma$(Si K$\alpha$)&I(\MgXII\ Ly$\beta$)&
$\sigma$(\MgXII\ Ly$\beta$)&$M_{BH}$ & $v_{FWHM}$ & r \\
& 10$^{-6}$ \# cm$^{-2}$\ s$^{-1}$ & eV & 10$^{-6}$ \# cm$^{-2}$\ s$^{-1}$ & eV &
$M_\odot$ & km\ s$^{-1}$ & pc \\
\hline
Circinus & $3.7\pm0.7$ &$1.4\pm0.6$ & $2.4\pm0.6$ &$1.3$ & $1.7\times10^6$& $570\pm240$
					&$0.03^{+0.06}_{-0.015}$ \\
Mrk 3    & $3.2\pm1.0$ &$1.8\pm0.8$ & $2.3\pm0.9$ &$3.4$ & $4.5\times10^8$& $730\pm320$
					&$4.9^{+9.7}_{-2.7}$ \\
NGC 1068 & $6.4\pm1.5$ &$0.8\pm0.7$ & $3.5\pm1.2$ &$2.0$ & $8.6\times10^6$& $320\pm280$
					&$0.5^{+30}_{-0.36}$\\
\hline
\end{tabular}
\begin{description}
\begin{footnotesize}
\item
  Note: I is the fitted line intensity and $\sigma$ the fitted line width. 
  $\sigma$(\MgXII\ Ly$\beta$) is fixed at the value measured 
  from the \MgXII\ \Lya line. The masses of the black hole ($M_{BH}$) are quoted from \citet{Gre03},
  \citet{LB03}, and \citet{WU02}, respectively. 
\end{footnotesize}
\end{description}
\end{table*}

\section{Observational data }

One complexity of using the Si \Ka line (1.74 keV) is that
it is blended with the \MgXII\ \Lyb line at 1.745 keV (the fluorescent wavelengths 
are from Bearden 1967, while the other wavelengths are from NIST\footnote{www.nist.gov/pml}).
The peaks of these two lines can be resolved by HEG, provided that the signal
is good enough.
We searched the \cha data archive for Compton-thick AGN observed with HETGS and
found that the data of the Circinus galaxy, Mrk 3, and NGC 1068
can be used to measure the two lines simultaneously. 
The exposure times are 600, 400, and 400 ks, respectively.
We use the data downloaded from \cha Transmission Grating Data Archive and Catalog 
\citep[TGCat,][]{TG}.
All the spectra are extracted from a region with a 2 arcsec half-width in the cross-dispersion
direction.
The data from $\pm1$ order of HEG are combined together.
The HEG spectra of all the three sources have been published in the literature,
such as \citet{Sam01} and \citet{Are14} for Circinus galaxy, \citet{Sak00} for Mrk 3, 
and \citet{Ogl03} and \citet{Kal14} for NGC 1068.

\begin{figure}
\includegraphics[width=2.65in]{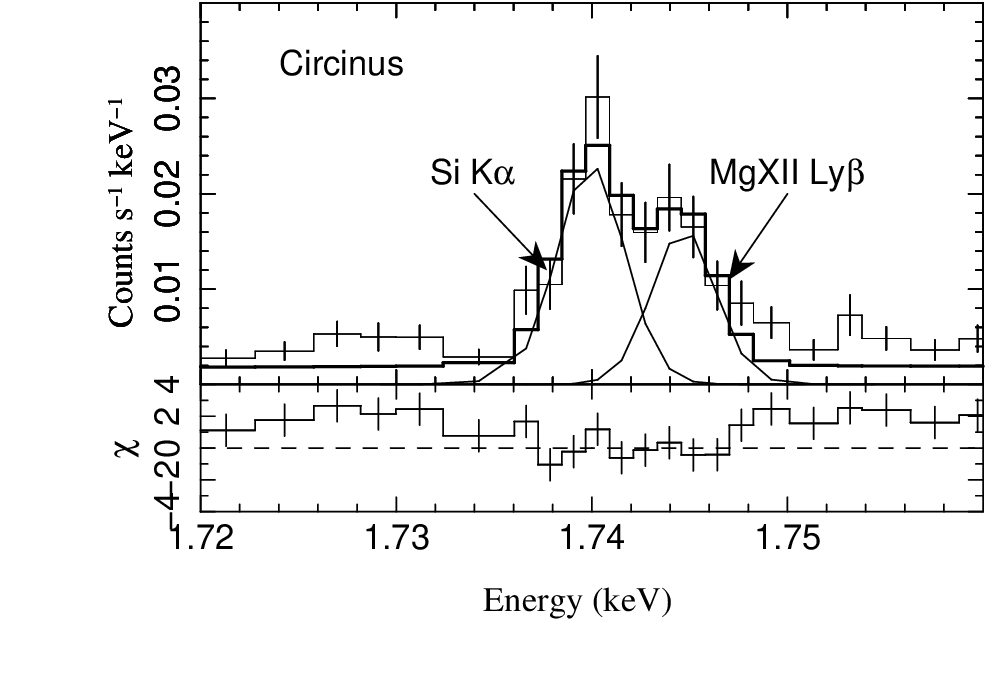}
\includegraphics[width=2.65in]{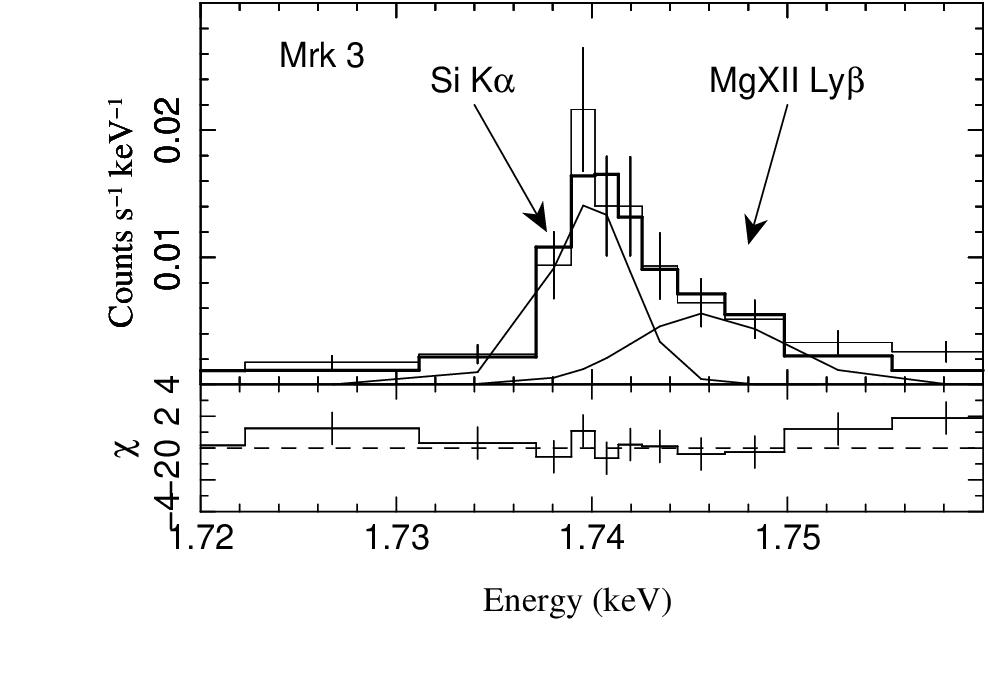}
\includegraphics[width=2.65in]{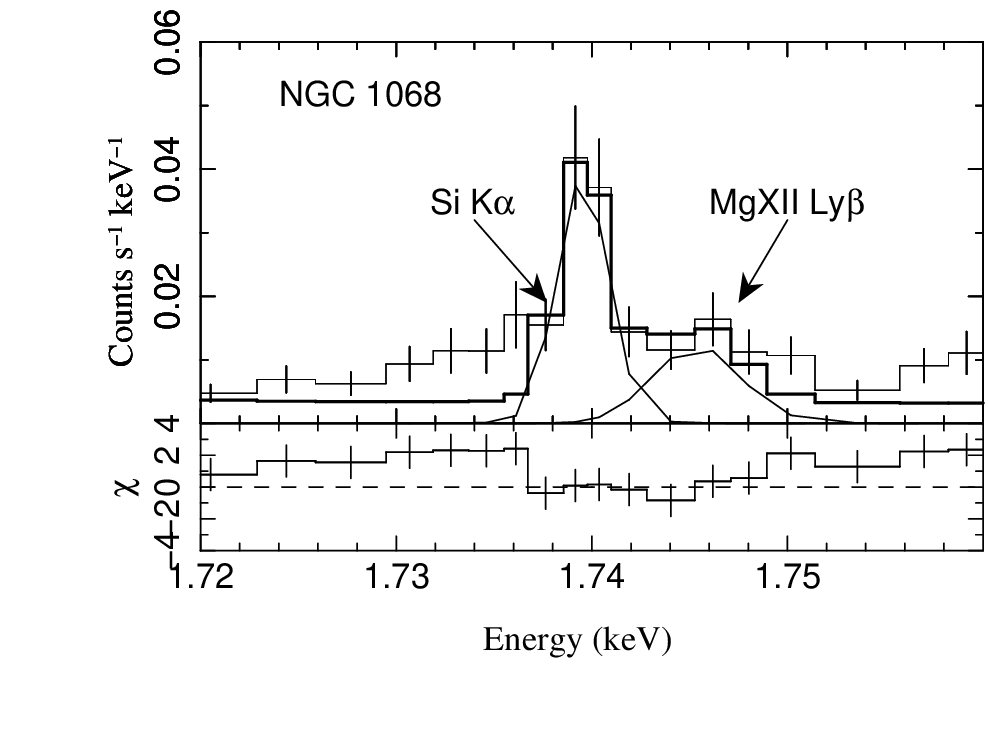}
\caption{
The \cha HEG spectra of the Si \Ka and \MgXII\ \Lyb line complex
of Circinus, Mrk 3, and NGC 1068, corrected for redshift.
 The fitted models of two Gaussian lines plus a power-law continuum 
are plotted as the thick solid lines, while the thin lines indicate individual components.
}
\end{figure}

\section{Results}

The line complexes around 1.74 keV of Circinus, Mrk 3, and NGC 1068 observed with \cha HEG
are plotted in Fig. 1. The data are binned to 0.005 \A with a minimum signal-to-noise ratio of 3.3. 
As can be seen, the line complexes are dominated by the 
Si \Ka line at 1.74 keV, but skewed toward higher energy where the \MgXII\ \Lyb line
is located. The peaks of the two lines
 are resolved for Circinus and NGC 1068, while they are blended for Mrk 3.
To check the degrading of the spectral resolution due to the spatial extent of the Si \Ka line, 
we extracted the radial profile of the 1.74 keV line complex (with a width of 20 eV)
from the zero-order images of the three sources, and found that 
it is consistent with the instrument PSF profile for all three sources.
Therefor, the HEG resolving power is not affected much by the spatial extent 
of the Si \Ka line.

We fit the line complex with two Gaussian lines plus a power-law continuum.
The shift and width of the \MgXII\ \Lyb line are fixed with the values measured from 
the \MgXII\ \Lya line at 1.472 keV. Since the \MgXII\ \Lya and \Lyb lines are from 
the same ions, their line properties should be the same.
The fitting region is between 1.5 and 2.3 keV by neglecting the other main emission features. 
The fitted results are plotted in Fig. 1 and listed in Table 1.

From Fig. 1 we see that 
the line complexes around 1.74 keV are reasonably fitted by the adopted
model.
The residuals around 1.73 keV are likely due to the \AlXIII\ \Lya line, while 
those above 1.75 keV are likely due to the fluorescent lines from singly ionized Si$^{+}$
\citep[1.753 keV,][]{Bea67}, the fluxes of which are much smaller compared with those from 
neutral Si. The fitted line widths of the Si \Ka line are $1.4\pm0.6$, 
$1.8\pm0.8$, and $0.8\pm0.7$ eV,
for Circinus, Mrk 3, and NGC 1068, respectively.
They correspond to FWHM velocities ($v_{FWHM}$) of $570\pm240$, $730\pm320$, and $320\pm280$ 
km\ s$^{-1}$.
Assuming the line-emitting material is in a virialized orbit around the black hole,
its distance to the black hole can be estimated as \citep[][]{Net90} 
$r=4c^2/3v_{FWHM}^2r_g$, where $r_g=GM_{BH}/c^2$.
Adopting the black hole masses listed in Table 1, the distances of the line-emitting
region are 0.03 pc, 4.9 pc, and 0.5 pc, for Circinus, Mrk 3, and NGC 1068, respectively.
We note that the numerical factor $4/3$ in the above formula suffers from the 
uncertainty of the geometrical and dynamical assumptions of the line-emitting material.
For example, if it is in a shape of thick disk with $h/r<1$, the factor becomes 
$4(h^2/r^2+\sin^2{\it i})$, where ${\it i}$ is the inclination to the line of sight \citep{Net13}.
On the other hand, if the line-emitting material is in some kind of outflow, the factor 
will depend on the dynamical modelling of the flow.

\section{Conclusion and Discussion}

X-ray fluorescent lines are unique features of the reflection spectrum 
emitted by the torus when irradiated by the central AGN. They are most
prominent for Compton-thick sources, the intrinsic spectra of which are 
hidden from direct viewing. The intrinsic line width of the X-ray fluorescent 
lines can be used to probe the location of the line-emitting region. 
Using the Si \Ka line at 1.74 keV, for which the \cha HEG 
has a spectral resolution $\sim4$ times better than for the Fe \Ka line at 6.4 keV,
we measured the FWHM of the Si \Ka line for Circinus, Mrk 3, and NGC 1068.
They have a FWHM of 570, 730, 320 km\ s$^{-1}$, respectively. These values are $3-5$ times
smaller than those measured with the Fe \Ka line,
which are 1710, 3140, and 2660 km\ s$^{-1}$, respectively \citep[][]{Shu11}.

While the ionization potentials of the valence electrons of neutral Si and Fe 
are similar, the ionization potential of the K-shell electrons of Fe is larger
than that of Si. Since the hard photons can penetrate deeper into the material,
in principle, the emitting region of the Fe \Ka line should not be smaller than 
that of the Si \Ka line. Therefore the intrinsic line widths of the Fe \Ka line
should not be larger than those of the Si \Ka line, and they
are most likely to be over-estimated. While the limited spectral resolution is likely to be 
a major reason for the over-estimation, there are several other reasons that 
could lead to a large measured width of the Fe \Ka line.

First, the neutral Fe \Ka line is composed of a doublet, K$\alpha_1$ at 6.404 keV,
and K$\alpha_2$ at 6.391 keV, with a flux ratio of 2:1 \citep{Bea67}.
This leads to a difference of 13 eV (610 km\ s$^{-1}$), which is comparable to the FWHM of HEG  
at 6.4 keV. Secondly, the observed Fe \Ka line may not come only from the neutral
Fe, and is likely blended with those from other low-ionized Fe ions, the energies of which 
are higher than 6.4 keV by tens of eV \citep{KM93}. 
Thirdly, the Compton scattering of the Fe \Ka photons also tends to enlarge the width 
of the Fe \Ka line. These intrinsic spreading makes the measurement of the intrinsic 
line width of the Fe \Ka line difficult even with instruments of higher spectral resolution.
Therefore one should be cautious when using the measured width of the Fe \Ka line to infer the
line-emitting region.

The Si \Ka line is also composed of a doublet (at 1.7400 keV and 
1.7394 keV) with a difference of 0.6 eV (100 km\ s$^{-1}$), which is well below 
the HEG spectral resolution. 
As the Si \Ka line is well resolved from the singly ionized Si$^{+}$ \Ka line (1.753 keV),
it represents the truly cold torus, different from the Fe \Ka line, which may be contaminated
by ionized Fe ions.
It is interesting to compare the estimated distances of the line-emitting region
with the dust sublimation radius $r_{dust}$. Using the scaling relation obtained from 
the reverberation measurements of the inner dust radius by \citet{Kos14}, 
the $r_{dust}$ are 0.02, 0.15, and 0.11 pc for Circinus, Mrk 3, and NGC 1068, respectively.
Therefore, the estimated line-emitting regions are located outside the $r_{dust}$.
This is in contrast to the recent claims that the line-emitting regions are 
within the $r_{dust}$ based on the measured width of the Fe \Ka line \citep[e.g.][]{MM15,Gan15}.
The line-emitting distances estimated from the width of Si \Ka line are 
also consistent with the recent infrared 
interferometric observation of Circinus on sub-parsec scale \citep{Tri14} 
and NGC 1068 on parsec scale \citep{Lop14}.
It shows that for Compton-thick AGN, the X-ray fluorescence material are likely 
to be the same as the dusty torus emitting in the infrared.

The samples we studied only include 3 galaxies. There are other several galaxies, for which
the Si \Ka lines are detected significantly, although the existing data are not deep enough 
to allow a deblending of the \MgXII\ \Lyb line yet, such as NGC 4507 and NGC 7582.
With further observations of moderate exposure times ($200-300$ ks), the kind of studies conducted
here can also be applied to these samples.
Beside the Si \Ka line, the fluorescent \Ka lines of S, Ar, and Ca are also detected 
for all three sources, but with a lower S/N due to the quick decline of the effective 
area of HEG above 2 keV. The signals are generally too weak to provide useful constraints
on the line width. Nevertheless, the effective area around these lines will be improved 
with the forthcoming soft X-ray calorimeter on-board Astro-H, which has a spectral 
resolution $\sim7$ eV.

Because the Si \Ka line is more sensitive to the absorption than the Fe \Ka line,
the flux ratio between the Fe \Ka line and Si \Ka line will be a useful probe of the torus. 
For example, if the torus is clumpy, the Si \Ka line will not be significantly reduced,
due to its pass through the free-of-obscuration regions between the clumps 
\citep[e.g.][]{NG94,Liu14}.
Unlike the measurement of the line widths presented here, which needs deep data to resolve
the \MgXII\ \Lyb line, the Fe K$\alpha$/Si \Ka ratio can be calculated for a larger sample with less S/N.
We have compared the Fe K$\alpha$/Si \Ka ratio of 10 type II AGN with the simulation of clumpy 
torus \citep{Liu14}, and found that the clumpy torus is preferred. The detailed results are presented 
in \citet{Liu16}. These results show that non-Fe fluorescent lines at soft X-ray band are a potential
powerful probe of the torus around AGN.

\section*{Acknowledgements}
We thank our referee for constructive suggestions and Liu Yuan for valuable comments on an early draft. 
JL is supported by NSFC grant 11203032.
This research is based on data obtained from the \cha Data Archive.

\bibliographystyle{mn2e}

\end{document}